\documentclass[12pt]{article}
\usepackage{amsmath,amssymb,amsfonts}
\usepackage{bbm}
\usepackage{graphicx}
\usepackage{color}
\begin{document}
\author{Yan V. Fyodorov\\ \textit{Department of Mathematics, Kings College London, London WC2R 2LS, UK}\\[0.25cm]
  Sven Gnutzmann\\ \textit{School of Mathematical Sciences, University of Nottingham, Nottingham NG7 2RD, UK}\\[0.25cm]
  Jonathan P. Keating\\ \textit{School of Mathematics, University of Bristol, Bristol BS8 1TW, UK}}

\title{Extreme values of CUE characteristic polynomials: a numerical study}
\date{\today}

\maketitle
\abstract{We present the results of systematic numerical computations
  relating to the extreme value statistics of the characteristic
  polynomials of random unitary matrices drawn from the Circular
  Unitary Ensemble (CUE) of Random Matrix Theory.  In particular, we
  investigate a range of recent conjectures and theoretical results
  inspired by analogies with the theory of logarithmically-correlated
  Gaussian random fields.  These include phenomena related to the
  conjectured freezing transition.  Our numerical results are
  consistent with, and therefore support, the previous conjectures and
  theory.  We also go beyond previous investigations in several
  directions: we provide the first quantitative evidence in support of
  a correlation between extreme values of the characteristic
  polynomials and large gaps in the spectrum, we investigate the rate
  of convergence to the limiting formulae previously considered, and
  we extend the previous analysis of the CUE to the C$\beta$E which
  corresponds to allowing the degree of the eigenvalue repulsion to become a
  parameter.}
\section{Introduction and Background}

Questions related to quantifying statistical properties of high and
extreme values taken by the characteristic polynomials of random
matrices have recently attracted considerable
attention~\cite{ABB,CMN,FHK_prl,FyoKeat2014,FyoSimm,Keating_lec,Najnudel,PZ2017}.
The main motivation was the suggestion of a close
analogy~\cite{FHK_prl,FyoKeat2014} between the statistics of the
(logarithm of) the modulus of characteristic polynomials of large
random matrices~\cite{HKOC} (originally, from the Circular Unitary
Ensemble, or CUE, of Random Matrix Theory) and an important class of
log-correlated random processes and fields, namely those characterised
by having a logarithmic singularity on the diagonal of the covariance kernel,
which have been the focus of considerable attention in the past few
years.  Such processes and fields appear with surprising regularity in
many different contexts, ranging from the statistical mechanics of
branching random walks and polymers on trees~\cite{DS1988,CLD} and
disordered systems with multifractal structure~\cite{CMW,logmultlec}
to models of random surfaces underlying the probabilistic description
of two-dimensional gravity~\cite{RVLectures,KRV-DOZZ}.  In particular,
the extremal values of log-correlated Gaussian processes have been the
subject of intensive study, leading to considerable progress ranging from non-rigorous~\cite{FB2008,FLDR,FLD2016},
through tending-to-rigorous~\cite{Ostr2016,OstrRev}, to fully rigorous
\cite{DRZ,SubagZeitouni,Remy1,Remy2} analysis.

The characteristic polynomials of random matrices are of considerable
interest in their own right, but in addition they underpin an
influential model~\cite{KS2001} for the statistical properties of the
Riemann zeta function on the critical line~\cite{Keating_lec}.  One of
the main points of the articles~\cite{FHK_prl,FyoKeat2014} is that on
the global and mesoscopic scales, one should think of the logarithm of
the zeta function on the critical line as behaving statistically like
a log-correlated field. Large values of the Riemann zeta on the
critical line are of considerable and long-standing
interest~\cite{Keating_lec}, and this new perspective has
attracted a good deal of attention
~\cite{ABH,ABBRS,Najnudel,SaksWebb}. 

It is of course important to make
clear that every step of the translation should be taken with
appropriate adjustment and caution. Log-correlated fields are
necessarily random generalised functions (distributions), and
establishing in which way the logarithm of the modulus of a
characteristic polynomial tends to such a highly singular object is a
non-trivial task (see, for example,~\cite{HKOC}). Such studies
necessarily involve regularisations, and one of the most natural
rigorous frameworks seems to be provided by the theory of multiplicative
chaos, which has its origins in informal ideas of Mandelbrot
\cite{Mand} which were developed into a comprehensive, mathematically rigorous theory
in works by Kahane~\cite{Kahane1}, and have been further
extended in recent years~\cite{BerLectures,RVLectures}.  The
convergence of characteristic polynomials and the Riemann
zeta-function on the critical line, at the appropriate scale, to the
Multiplicative Chaos measures was established in several recent
papers~\cite{BWW,Webb}.  Together with other related results,
e.g.~\cite{FyoKhorSimm,PZ2017} this provides strong support for the
correspondence in question.

Despite these developments, up to now only a few preliminary attempts
have been made to investigate the predictions of the theory by direct
numerical simulations of large random
matrices~\cite{FHK_prl,FyoKeat2014,FyoSimm,Keating_lec}.  In this
paper we present the results of an extensive numerical investigation
of large CUE matrices when the matrix dimension $N=2^M$ is large.  In
our experiments we diagonalised numerically more than $10^7$ matrices
for $M=2,3,4,5,6,7,8,9$ and somewhat fewer matrices for $M=10,11$ and
$M=12$ (more than $10^6$ for $M=10$, more than $10^5$ for $M=11$, and
$50000$ for $M=12$), and extracted the relevant information from their
spectra.

Our first goal is to check manifestations of the main mechanism
underlying the extreme value statistics of log-correlated processes,
the so-called 'freezing transition'~\cite{CLD,FB2008,SubagZeitouni}.
In fact, some features of this phenomenon are seen already at the
level of sequences of {\it independent} (rather-than log-correlated)
random sequences.  Indeed, it was first discovered at the level of the
so-called uncorrelated Random Energy Model (REM) introduced originally
by Derrida~\cite{Derrida_REM_prb}, which is a toy model for such a
freezing transition.  In REM the energy values $\{E_n\}_{n=1}^N$, are
taken to be i.i.d. random variables with a Gaussian distribution
\begin{equation}\label{REM}
  P(E)= \frac{1}{\sqrt{M \pi J^2}} e^{- \frac{E^2}{M J^2}}, \quad M=\log{N}/\log{2}
\end{equation}
where $J>0$ is a global energy scale. Defining for a temperature $T>0$
the REM partition function as ${\cal Z}_N=\sum_{n=1}^N e^{-E_n/T}$ one
then can study the associated free energy ${\cal F}=-M^{-1}T\log{{\cal
    Z}_N}$, and show that its mean value becomes independent of $T$
below some finite critical temperature $T=T_c$.  

Although the
log-correlated models share the existence of a freezing transition
with the simple REM, finer features implied by freezing are quite
different, and the two models actually belong to different
universality classes.  The difference is reflected, in particular, in
much broader fluctuations in the number of points in the process
exceeding a high threshold in the log-correlated case~\cite{FLDR2},
and eventually in the statistics of extreme values.  From that angle,
we always compare the results for CUE simulations with the
corresponding numerical simulation of the simple REM
at comparable ensemble sizes (and beyond).  Our investigations shed
some light on the convergence properties of the extremal value
distribution, the distribution of partition function moments, and the
associated mean free energy.

Arguably, the most well studied, and, in a sense, paradigmatic example
of a $1D$ processes with logarithmic correlations is the so-called
Gaussian circular-logarithmic model suggested originally
in~\cite{FB2008}.  The process has a simple representation as a
(formal) Fourier series given by the real part of $\sum_{n=1}^{\infty}
v_n \frac{e^{-i n t}}{\sqrt{n}}$, with i.i.d. complex random Gaussian
coefficients $v_n$ with zero mean and unit variance.  The same object
can be, with due interpretation, viewed as a 2D Gaussian free field
defined on a disc with Neumann boundary conditions and sampled along a
circle of unit radius, see \cite{AJKS,Remy1}.  As such, it has
intimate connections with conformal field theory, and appears in that
context very naturally~\cite{AJKS}. This line of research very
recently provided the first rigorous proof~\cite{Remy1} of the
conjectured explicit distribution of its extreme values, see
Eq.~(\ref{FBdistr}) below.  Fortunately, it is precisely the model
which is expected to represent the limiting statistics of the log-mod
of CUE characteristic polynomial, so will serve as a benchmark for our
numerics.

Let us finally mention that, given the widespread interest in the family of 
$\beta-$ensembles of symmetric three-diagonal random
matrices\footnote{The Dyson parameter $\beta$ in the definition of
  the $\beta$-ensembles should not be confused with the notation for
  the inverse temperature $\beta=1/T$ used elsewhere in the paper,
  apart from the Appendix B.} introduced by Dumitriu and Edelman
\cite{DumEdel} and their {\it circular} analogues (C$\beta$E)
\cite{killip}, it seems natural to ask if the large-$N$ statistics of
the extreme values for characteristic polynomials in this family will
have similar properties to the $\beta=2$ case for any fixed $\beta>0$.
The first steps in this direction were taken by Chhaibi et
al.~\cite{CMN} who proved that, up to a simple rescaling of
parameters, the first two (non-random) terms in the asymptotics of the
maximum value, see Eq.(\ref{shiftmax}), are indeed common to all
members of the family.  In Appendix B we extend to arbitrary
$\beta>0$ heuristic computations given in \cite{FyoKeat2014} for the
CUE.  On this basis we conjecture that the distribution
Eq.(\ref{FBdistr}) of the first nontrivial random term should be also
universal with respect to changes of the parameter $\beta$.

The structure of the paper is as follows. We introduce some notation
and definitions in subsection \ref{sec:setting}.  We present our
numerical results for the extreme value statistics in section
\ref{sec:extreme_values}.  In section \ref{sec:free} we present data
supporting the conjectured freezing transition for the free energy.
We summarise our conclusions and outlook in section
\ref{sec:conclusions}.  The paper has two appendices.  In the first
appendix we compare our numerical data with some recent exact formulae
for the moments of the partition function.  In the second appendix we
present a heuristic calculation for the C$\beta$E which generalises
that given in \cite{FyoKeat2014} for the CUE.

{\bf Acknowledgements}.  The research at King's College (YF) was
supported by EPSRC grant EP/N009436/1 "The many faces of random
characteristic polynomials".  JPK is grateful for support from a Royal Society Wolfson Research Merit Award and ERC Advanced Grant 740900 (LogCorRM).

\subsection{Setting: extreme values of the characteristic polynomial,
  moments and free energy for CUE and REM}
\label{sec:setting}

Let $U_N \in U(N)$ be a unitary $N \times N$ matrix taken at random
from the CUE, i.e. uniformly with respect to the Haar measure on
$U(N)$. Denoting its eigenvalues by $\{e^{i \phi_n}\}_{n=1}^N$, its
characteristic polynomial is given by
\begin{equation}\label{defcharpol}
  p_N(\theta)= \det(1-U_N e^{-i \theta})= \prod_{n=1}^N
  (1-e^{i(\phi_n-\theta)}) .
\end{equation}
We will be interested in the statistical distribution of $|p_N(\theta)|$ with a focus on its extremal (maximal) value
\begin{equation}
  |p_N|_{\mathrm{max}}\equiv \mathrm{max}_{\theta \in [0,2\pi)}\ |p_N(\theta)|\ .
\end{equation}
Let us express the square of the maximum as $|p_N|^2_{\mathrm{max}} =
e^{-a_N+ b_N y}$ where $a_N$ and $b_N$ are coefficients that depend
only on the matrix size $N$ which will be discussed later in detail. The
limiting distribution, as $N\to \infty$ of the variable $y$ has been
in the centre of recent conjectures~\cite{FHK_prl,FyoKeat2014} that
will be described more fully in Sect.~\ref{sec:extreme_values}. 
For a detailed description of the technical background we
refer the reader to \cite{FyoKeat2014}. In the present paper we will just
describe the setting, the main definitions and the main conjectural predictions
without much detail how these were obtained.
We then focus on the detailed description of the comparison of
numerical data with these predictions. 

The distribution of absolute values $|p_N(\theta)|$ of the
characteristic polynomial and its extremal values may be characterised
in terms of the moments
\begin{equation}
  \mathcal{Z}_N(\beta)= \frac{N}{2\pi} \int_0^{2\pi} \left| p_N(\theta)
  \right|^{2\beta} \equiv \frac{N}{2\pi}\int_{0}^{2 \pi} e^{- \beta V_N(\theta)}
\end{equation}
where $V_N(\theta)=- 2 \log |p_N(\theta)|$. In analogy to the partition
function in statistical physics we refer to $\beta$ as the inverse
temperature and introduce the normalised free energy
\begin{equation}
  \mathcal{F}_N(\beta)= -\frac{1}{\beta \log N} \log
  \mathcal{Z}_N(\beta)\ .
\end{equation}
The additional normalisation with $\log N$ is to ensure the existence
of a finite limiting value as $N \to \infty$.  The extreme values are
obtained in the low temperature limit $\beta \to \infty$ of the free
energy via
\begin{equation}
  \lim_{\beta\to \infty} \mathcal{F}_N(\beta) = \frac{1}{\log N} \mathrm{min}_{\theta
    \in [0,2\pi)} V_N(\theta)= - \frac{1}{\log N} \log
  |p_N|^2_{\mathrm{max}} .
\end{equation}

Our numerical investigation focuses on the statistical properties of
$\log |p_N|_{\mathrm{max}}$ and the moments $\mathcal{Z}_N(\beta)$.
Both properties only depend on the spectrum of CUE matrices.
Numerically we used the fact that the eigenvalues of CUE matrices have
the same joint probability distribution as explicitly known ensembles
of banded (5-diagonal) unitary matrices \cite{killip} which are much
easier to construct than full CUE matrices.  We used standard
octave/matlab routines for random number generators and numerical
diagonalisation.

As mentioned above we compare our results to REM in order to show the
characteristic differences between the two universality classes (and
for benchmarking).  In REM we set the energy scale to
\begin{equation}
  J= 2 \sqrt{\log 2}
\end{equation}
throughout this manuscript.  This choice ensures that the critical
inverse temperature for the freezing transition in REM is
$\beta_{\mathrm{crit}}=1$ in coincidence with the
critical inverse temperature in CUE.\\
We will compare the statistics of extreme values of the characteristic
polynomial of random unitary matrices in the form $- 2 \log
|p_N|_{\mathrm{max}}$ to the statistics of ground state energies
$E_{\mathrm{min}}=\mathrm{min}_{n=1}^N E_n$ for realisations of the
REM (due to the symmetry of the model this is equivalent to comparing
the maximal energy with $2\log|p_N|_{\mathrm{max}}$ ).  For
convenience (in order to have the same notation for CUE and REM) we
set $\log |p_N|_{\mathrm{max}}=-E_{\mathrm{min}}/2$
when considering REM.\\
Analogously we compare moments $\mathcal{Z}_N(\beta)$ for CUE with
partition sums
\begin{equation}
  \mathcal{Z}^{\mathrm{REM}}_N(\beta)= \sum_{n=1}^{N=2^M} e^{- \beta
    E_n}
\end{equation}
with the normalised free energy\footnote{Note that our normalisation
  with $\log N$ corresponds directly to the one used for CUE but
  differs by a factor $\log 2$ from the one used in most literature on
  REM (where $\log N$ is often replaced by $M=\frac{\log N}{\log
    2}$).}
\begin{equation}
  \mathcal{F}^{\mathrm{REM}}_N(\beta)= -\frac{1}{\beta \log N} \log
  \mathcal{Z}^{\mathrm{REM}}_N(\beta)\ .
\end{equation}
Our choice of parameters and normalisation implies that the theoretical 
prediction for the expected free energy in CUE and REM follow the same curve
\begin{equation}
  - \lim_{N\to \infty} \mathcal{F}^{\mathrm{CUE/REM}}_N(\beta)
  = 
  \begin{cases}
    \beta + \frac{1}{\beta} & \text{for $\beta\le 1$,}\\
    2 & \text{for $\beta \ge  1$.}
  \end{cases}
\end{equation}

\section{The distribution of extreme values}
\label{sec:extreme_values}

The logarithm of the maximal values $|p_N|^2_\mathrm{max}$ for REM is
known to be distributed according to a Gumbel distribution, see
e.g. \cite{BM1997}.  To be more specific, after an appropriate
rescaling
\begin{equation}
  2 \log |p_N|_\mathrm{max}= -a_N + b_N y
  \label{rescaling}
\end{equation}
the integrated probability distribution for the random variable $y$
converges to
\begin{equation}
  I_{\mathrm{REM}}(y) \equiv
  I_{\mathrm{Gumbel}}(y)=\int_0^y P_{\mathrm{Gumbel}}(y')\ dy' =
  e^{-e^{-y}}\ .
\end{equation}

\begin{figure}[h!]
  \begin{center}
    \includegraphics[width=0.7\textwidth]{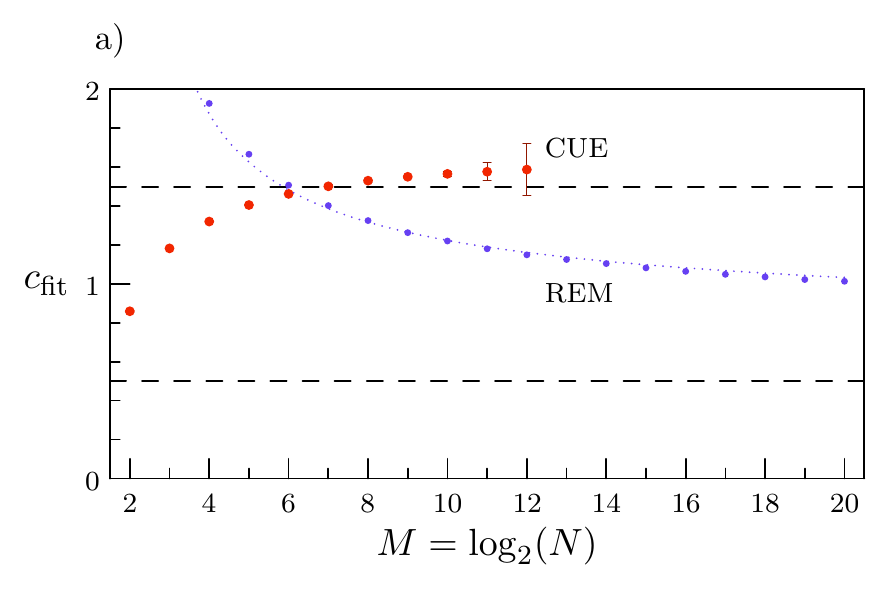}\\
    \hspace*{-0.8cm}\includegraphics[width=0.76\textwidth]{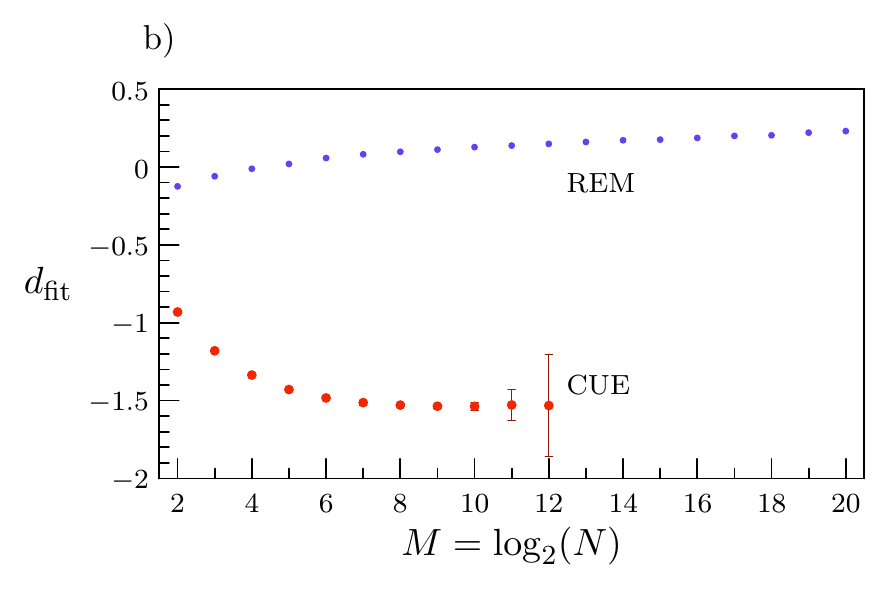}
  \end{center}
  \caption{\label{figure1} (Color online.) Scaling behaviour of the fitting parameters
    $c_{\mathrm{fit}}$ (upper panel) and $d_{\mathrm{fit}}$ (lower
    panel) with $N=2^M$ for CUE (large red dots) and REM (small blue dots) 
    data of
    $|p_N|_{\mathrm{max}}$. \newline Note that $\log_2 N=\log N/\log 2= M$ denotes
    the logarithm with base 2.\newline The error bars describe one
    standard deviation of the data (where no error bars are visible
    they are smaller than the dot size). The (blue) dotted curve in
    the upper panel is $0.5+1.4/\log \log N$ and shows that the REM
    data is consistent with $c_{\mathrm{fit}}=1/2 +
    O\left(1/\log \log N \right)$. For the CUE data the deviations of
    $c_{\mathrm{fit}}$ from $c=3/2$ are consistent with a much faster
    decay (at least for the given range of matrix sizes $N$).}
\end{figure}

\begin{figure}[h!]
  \begin{center}
    \includegraphics[width=0.38\textwidth]{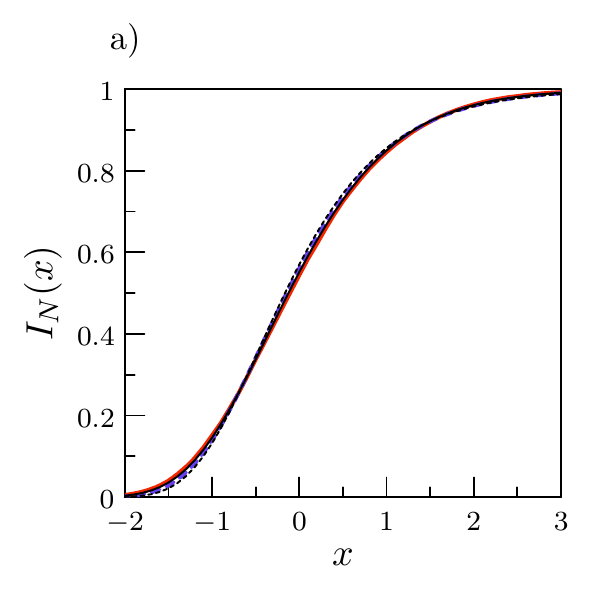}
    \includegraphics[width=0.25\textwidth]{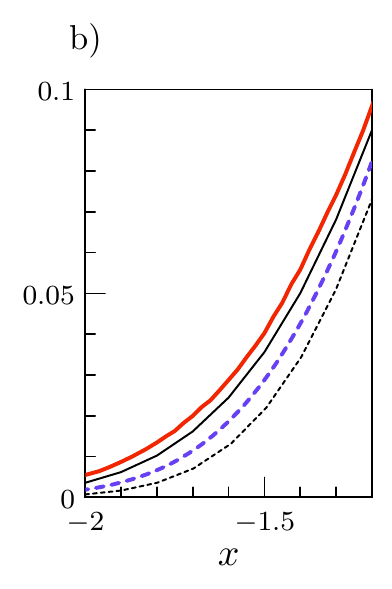}
    \includegraphics[width=0.25\textwidth]{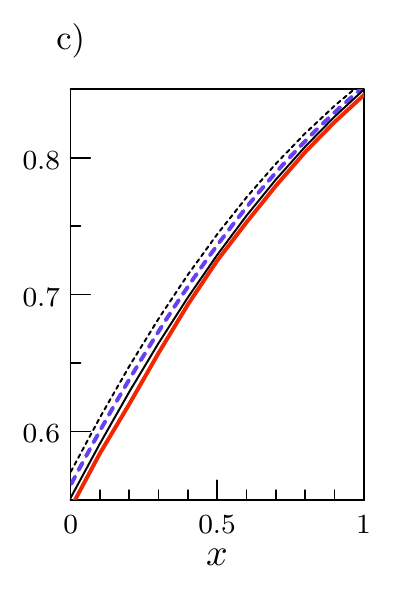}
  \end{center}
  \caption{\label{figure2} (Color online.) Integrated distribution function of
    $\log|p_N|_{\mathrm{max}}$ using the rescaled variable $x$ (see
    Eq.~\eqref{xscal}) for REM (blue dashed curve for $N=1048576=2^{20}$)
    and CUE (red curve for $N=4096=2^{12}$) against the predicted
    curves.\newline The full black line gives the CUE prediction. The
    dashed black line gives the Gumbel distribution (REM prediction).
    The left panel
    gives the whole distribution, the right two panels zoom into the
    distributions.  }
\end{figure}

\begin{figure}[h!]
  \begin{center}
    \includegraphics[width=0.4\textwidth]{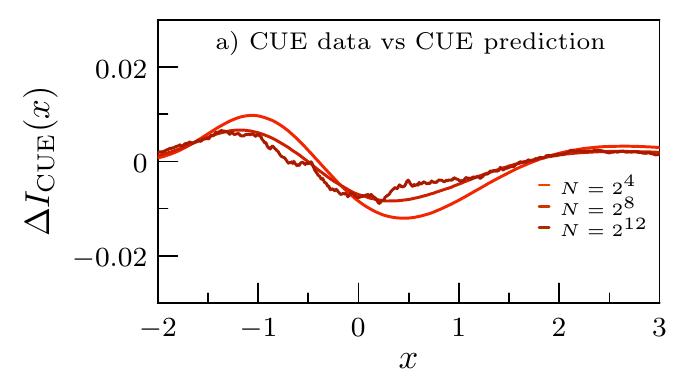}
    \includegraphics[width=0.4\textwidth]{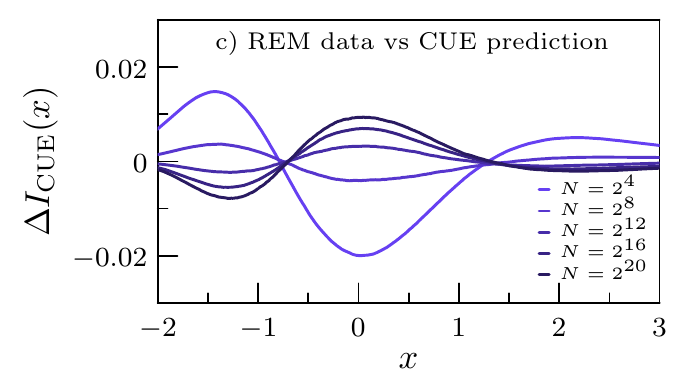}\\
    \includegraphics[width=0.4\textwidth]{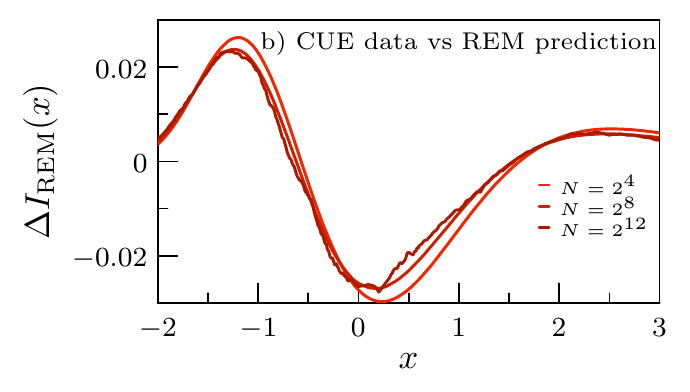}
    \includegraphics[width=0.4\textwidth]{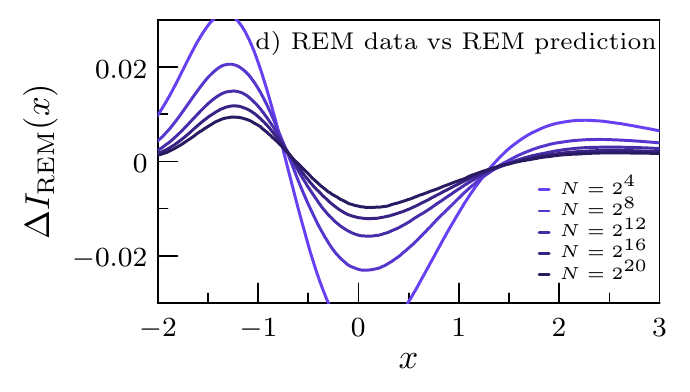}
  \end{center}
  \caption{
    \label{figure3}
    (Color online.)
    Difference between predicted integrated density distributions for
    $\log |p_N|_{\mathrm{max}}$ and numerically obtained distributions
    for various values of the size $N$ of the eigenvalue spectrum.
    The data is rescaled to vanishing mean and unit variance using the
    variable $x$ (see Eq.~\eqref{xscal}).\newline The left panels (a
    and b) show the difference between the two predictions and CUE
    data (for $N=2^4$, $2^8$, and $2^{12}$).\newline The right panels
    (c and d) show the difference between the two predictions and REM
    data (for $N=2^4$, $2^8$, $2^{12}$, $2^{16}$, and $2^{20}$).  }
\end{figure}

In the case of the CUE, appropriate rescaling leads to an integrated
distribution \cite{FB2008,FHK_prl,FyoKeat2014}
\begin{equation}
  \label{FBdistr}
  I_{\mathrm{CUE}}(y)= \int_0^y P_{\mathrm{CUE}}(y')\ dy' =
  2 e^{-y/2} K_1(2e^{-y/2})
\end{equation}
where $K_1(x)$ is the modified Bessel function of second kind and
order one.  
The Gumbel and the CUE distributions are
related by a simple convolution
\begin{equation}
  P_{\mathrm{CUE}}(y)=\int_{-\infty}^{\infty} P_{\mathrm{Gumbel}}(y_1)
  P_{\mathrm{Gumbel}}(y-y_1)\ dy_1.
\end{equation} 
In other words if $y_1$ and $y_2$
are two independent Gumbel-distributed random numbers then their sum
$y=y_1+y_2$ follows the CUE distribution.
The scaling parameters obey
\begin{equation}\label{shiftmax}
  \begin{split}
    a_N=&- 2 \log N + c \log \log N +o(1)\\
    b_N=& 1+ O(1/\log N)
  \end{split}
\end{equation}
in both cases. The constant $c$ however takes different values
\begin{equation}
  c=\begin{cases}
    \frac{1}{2} & \text{for $\mathrm{REM}$;}\\
    \frac{3}{2} & \text{for $\mathrm{CUE}$.}
  \end{cases}
  \label{eq:scalparam_c}
\end{equation}
The different value of this constant is a key signature of the long
range correlations in the CUE model.  We tested this numerically by
rescaling the data according to \eqref{rescaling} using $a_N\equiv
a_{\mathrm{fit}}$ and $b_N\equiv b_{\mathrm{fit}}$ as fitting
parameters such that the rescaled data has the same mean and variance
as the Gumbel distribution (for REM) or the CUE distribution (for
CUE).  From the fitted parameters we evaluated the quantities
\begin{equation}
  \begin{split}
    c_{\mathrm{fit}}=& \frac{a_{\mathrm{fit}}+2\log{N}}{\log  \log N}\ ,\\
    d_{\mathrm{fit}}=& (b_{\mathrm{fit}}-1) \log N\ .
  \end{split}
\end{equation}
From \eqref{shiftmax} we see that $c_{\mathrm{fit}}$ should
converge to $c$ with corrections of order $o(1/\log \log N)$ while
$d_{\mathrm{fit}}$ should be of order $O(1)$.  In Fig.~\ref{figure1}
we plot $c_{\mathrm{fit}}$ and $d_{\mathrm{fit}}$.  The plots are
consistent with the expected behaviour. Indeed, allowing deviations of
order $O(1/\log \log N )$ the data is consistent with
$c_{\mathrm{fit}} \to 3/2$ for CUE and $c_{\mathrm{fit}}\to 1/2$ for
REM (the difference between $O(1/\log \log N)$ and $o(1/\log \log
N)$ is too delicate to be resolved numerically).

\begin{figure}[htb]
  \begin{center}
    \includegraphics[width=0.7\textwidth]{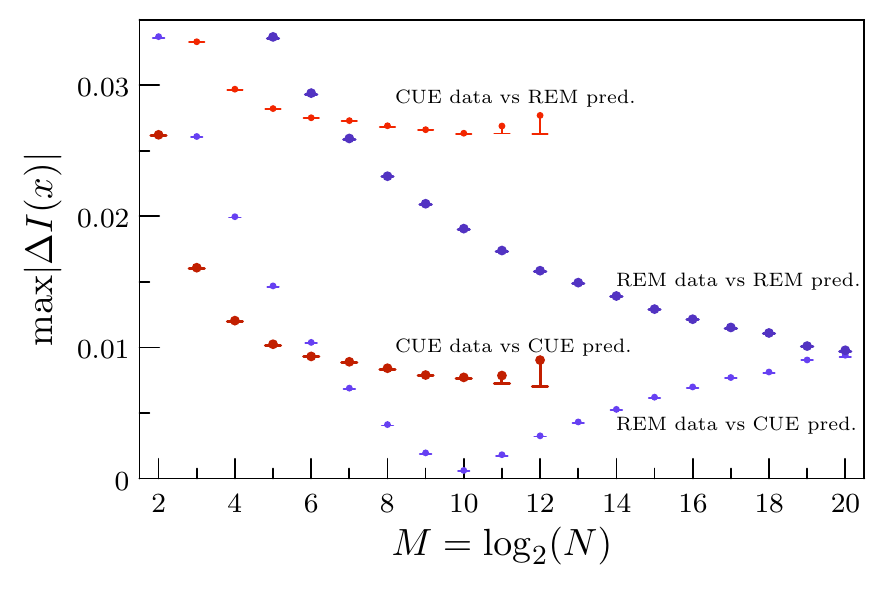}
  \end{center}
  \caption{\label{figure4} (Color online.) Maximal difference of the three predicted
    integrated density distributions against the data for CUE and REM
    as a function of size $N$ of the eigenvalue spectrum.  The
    one-sided error bars reflect the noise level -- the upper dot is
    pure data without any smoothing the lower bar gives the difference
    for smoothed data (local average over an interval of size
    $0.5$).\newline The thick (red) dots for $M=2$ to $M=12$ compare CUE
    data to the CUE prediction. The thin (red) dots for $M=2$ to $M=12$
    compare the CUE data to the REM prediction. The thick (blue) dots
    for $M=2$ to $M=20$ compare REM data to the REM prediction.  The thin (blue) dots
    for $M=2$ to $M=20$ compare REM data to the CUE prediction}
\end{figure}

A more detailed analysis of the distribution may be obtained by
comparing the integrated distribution functions directly. For this we
use the rescaled variable
\begin{equation}
  x = \frac{\log|p_N|^2_\mathrm{max} -
    \mathbb{E}[\log|p_N|^2_\mathrm{max}]}{
    \sqrt{\mathrm{Var}[\log|p_N|^2_\mathrm{max}]}
  }
  \label{xscal}
\end{equation}
which has vanishing mean and unit variance.  In Fig.~\ref{figure2} we
plot the integrated distribution from the REM and CUE data against the
predicted REM (Gumbel) and CUE distributions.  The latter have been
rescaled accordingly to have vanishing mean and unit variance and, with
minor abuse of notation, we will write
\begin{equation}
  I_{\mathrm{REM/CUE}}(x) \equiv
  I_{\mathrm{REM/CUE}}(y=\sqrt{\mathrm{Var}[y]}\ x+ \mathbb{E}[y] )\ .
\end{equation}
One can see in
Fig.~\ref{figure2} that the predicted curves and the curves obtained
from numerical data are all very close to each other. Zooming into the
details of the curves reveals that neither the REM curve has fully
converged to the REM prediction (at $N=2^{20}$) nor has the CUE curve
converged to the CUE prediction (at $N=2^{12}$).

We have analysed the convergence of the integrated distributions by
considering the differences
\begin{equation}
  \Delta I_{\mathrm{CUE/REM}}(x)= I_N(x) - I_{\mathrm{CUE/REM}}(x)
\end{equation}
of the data ($I_N(x)$) for REM and CUE and the three predicted
curves. In Fig.~\ref{figure3} we plot these differences for increasing
values of the size $N$ of the eigenvalue spectrum.  For REM the trend
in these differences is clearly consistent only with the REM
prediction. For CUE the picture is less clear because it was not
feasible to diagonalise matrices larger than $N=2^{12}$. We `only'
diagonalised 50000 matrices of that size $N=2^{12}$ which gives much
larger noise levels compared to smaller values of $N$ (at $N=2^8$ we
diagonalised $10^7$ matrices).  Nonetheless the numerics remains
consistent with the CUE prediction.

A more detailed analysis of the convergence may be obtained by
plotting the maximal difference $\mathrm{max}_{x \in \mathbb{R}}
|\Delta I_{\mathrm{CUE/REM}}(x)|$ (i.e. the $L^\infty$-norm of the
difference) against $N$, see Fig.~\ref{figure4}. This confirms again
that the REM data is consistent only with the REM prediction.
For CUE the available data indicates that the difference to the
REM prediction saturates at a finite value.  
The CUE data shows very slow convergence and the limited data
at $N=2048$ and $N=4096$ results in a large error term -- nonetheless
the data is consistent with convergence of the CUE data to the 
conjectured prediction for CUE.
Based on the data plotted in Fig.~\ref{figure4} one may
estimate that one may require $N \times N$ matrices with $N=2^{20}$ or
even larger in order to get a clearer support for the conjectured convergence.  
In order to get a sufficiently
smooth integrated distribution function one has to fully diagonalise
about $10^6$ to $10^7$ matrices. While there are specialised
algorithms for sparse or banded matrices that obtain a fraction of the
spectrum quite quickly we here need the full spectrum and computing
the required amount of data is well beyond our limits.

We would like to mention that the extreme value distribution for 
characteristic polynomials of the 
Gaussian Unitary Ensemble GUE ensemble has also been discussed 
recently~\cite{FyoSimm}.
GUE spectra have the same logarithmic correlations as CUE spectra and the 
corresponding distribution of extreme values of characteristic polynomials
shows similar deviations from the non-correlated REM spectra as CUE.
The predicted curve $I_{\mathrm{GUE}}(x)$ is different but very close to the
predicted curve $I_{\mathrm{CUE}}(x)$ 
(closer than to the REM curve $I_{\mathrm{REM}}(x)$). The origin of the difference is well
understood -- finite GUE spectra have Gaussian tails not present in CUE
(see for \cite{FyoSimm} for further details). We have checked whether our CUE
data is able to distinguish between the GUE and CUE predictions -- however 
the two are too close to be resolved.

\subsection{Correlations of the position of the maximal modulus of the
  characteristic polynomial and the spectrum}

\begin{figure}[htb]
  \begin{center}
    \includegraphics[width=0.7\textwidth]{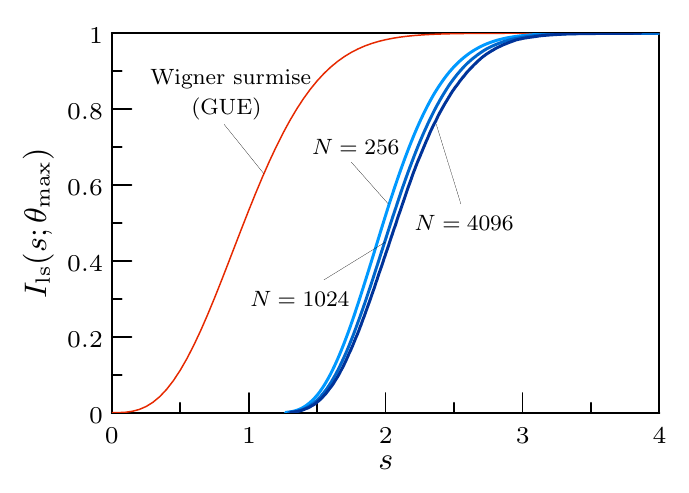}
  \end{center}
  \caption{\label{figure5} (Color online.)
    Integrated level spacing distributions
    $I_{\mathrm{ls}}(s;\theta_{\mathrm{max}})$ at
    $\theta_\mathrm{max}$ for CUE at $N=2^8=256$, $N=2^{10}=1024$ and
    $N=2^{12}=4096$. \newline The thin (red) curve on the left is the
    integrated Wigner surmise for unitary ensembles. The three thick
    (blue) curves give the CUE data.}
\end{figure}

\begin{figure}[htb]
  \begin{center}
    \includegraphics[width=0.65\textwidth]{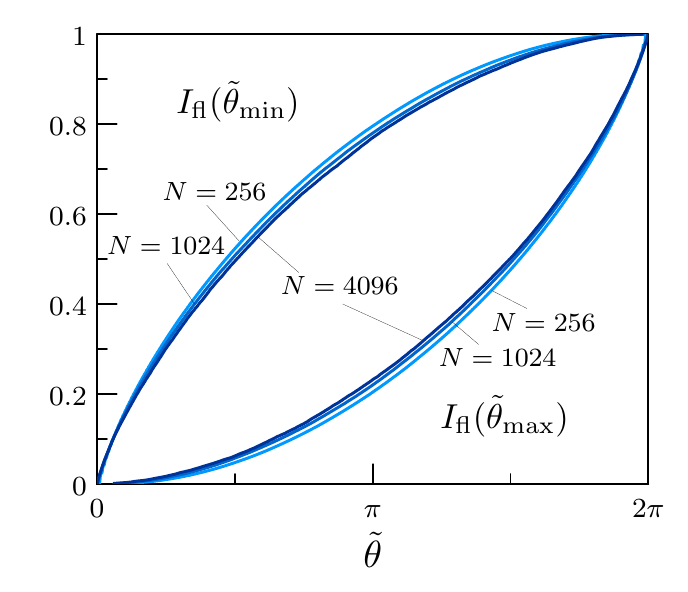}
  \end{center}
  \caption{\label{figure6} (Color online.) Integrated distribution
    $I_{\mathrm{fl}}(\tilde{\theta}_{\mathrm{max}})$
    ($I_{\mathrm{fl}}(\tilde{\theta}_{\mathrm{min}})$) of the position
    $\tilde{\theta}_{\mathrm{max}}$ ($\tilde{\theta}_{\mathrm{min}}$)
    where the fluctuations of the spectral counting function are
    maximal (minimal) measured from the position of the maximum of the
    modulus of the characteristic polynomial (at $\tilde{\theta}=0$).
  }
\end{figure}

The spectrum $\{e^{i \phi_n}\}_{n=1}^N $ of the unitary matrix $U$
gives the zeros of the characteristic polynomial $|p_N(\theta)|$. One
may expect that the position $\theta_{\mathrm{max}}$ where
$|p_N(\theta_{\mathrm{max}})|= |p_N|_{\mathrm{max}}$ statistically
occurs preferably in large intervals that are free of zeros.  This
kind of correlation may be measured in various ways. Most directly one
may consider the level spacing distribution at
$\theta=\theta_{\mathrm{max}}$.  
For any unitary matrix
of dimension $N$
the mean level spacing over its complete spectrum is (trivially)
$2\pi/N$. Numerically we find that typical values of level spacings
at $\theta_{\mathrm{max}}$ for CUE matrices are about twice as large
for the matrix sizes we used. This is a clear indication for
correlation between the position $\theta_{\mathrm{max}}$ and large
spacings for which we will give a more precise description in the
following. 
Indeed our numerical analysis allows for a more detailed
analysis of this deviation by considering the full statistical
distribution of level spacings at $\theta_{\mathrm{max}}$ for CUE
matrices.
While it is very hard to make strong
analytical predictions about the properties of this distribution (and we
are not aware of any relevant results) it is numerically
straightforward from the large amount of CUE spectra that we have
computed.  For each spectrum we have obtained the scaled levels
spacing $s=N \left(\phi_{n+1}-\phi_{n}\right)/(2\pi)$ where
$\phi_{n+1}>\theta_{\mathrm{max}}>\phi_{n}$.  The integrated level
spacing distribution $I_{\mathrm{ls}}(s; \theta_{\mathrm{max}})$ is
then the ratio of the number of spectra where the rescaled level
spacing is smaller than $s$ over the total number of spectra.  In
Fig.~\ref{figure5} we plot this quantity for $N=2^8$, $2^{10}$ and
$2^{12}$ and compare it to the Wigner surmise.  If
$\theta_{\mathrm{max}}$ was a typical point in the spectrum one would
expect to see a distribution close to the Wigner surmise and the
expectation value of the level spacing would be close to unity.  The
plots in Fig.~\ref{figure5} show however strong deviations from the
Wigner surmise and the expected level spacings are much larger than
unity (and growing slowly with $N$).  This is evidence for strong
correlations between the position $\theta_{\mathrm{max}}$ of the
maximum of the modulus of the characteristic polynomial and the
spectrum close to this value.

From the correlations between the position $\theta_\mathrm{max}$ and
the increased level spacings at this point one may expect further
correlations between the maximal values of the characteristic
polynomial and the spectrum.  Let
$\tilde{\theta}=\theta-\theta_{\mathrm{max}} \in [0,2\pi)$.  The
spectral counting function may be written as
\begin{equation}
  N(\tilde{\theta})=\frac{N \tilde{\theta}}{2\pi}
  - \frac{1}{\pi} \mathrm{Im}\, \log
  \det(1-e^{i \theta_{\mathrm{max}}}U)+ \frac{1}{\pi} \mathrm{Im}\, \log
  \det(1-e^{-i(\tilde{\theta}+\theta_{\mathrm{max}})}U) \ .
\end{equation}
In this form it counts the number of states above
$\theta_{\mathrm{max}}$ and it directly relates the spectrum of $U$ to
the characteristic polynomial.  The fluctuations in the spectral
counting function
\begin{equation}
  N_{\mathrm{fluct}}(\tilde{\theta})=\frac{1}{\pi} \mathrm{Im}\, \log
  \det(1-e^{-i(\tilde{\theta}+\theta_{\mathrm{max}})}U)
\end{equation}
are expressed in terms of the argument of the characteristic
polynomial $p_N(\theta)$. One may expect that the position of the
maximum is correlated with large fluctuations.  Numerically we obtain
the values $\tilde{\theta}_{\mathrm{max}}$ and
$\tilde{\theta}_{\mathrm{min}}$ where
$N_{\mathrm{fluct}}(\tilde{\theta})$ takes its maximal and minimal
values.  In Fig.~\ref{figure6} we plot the integrated density
$I_{\mathrm{fl}}(\tilde{\theta})$ of these values as a function of
$\tilde{\theta}$.  In absence of correlations one expects a straight
line $\tilde{\theta}/(2 \pi)$.  However, the plots show deviations
from a straight line that imply strong correlations between the
position of the maximum of the modulus of the characteristic
polynomial and the positions of the extrema in the fluctuations of the
spectral counting function.  The plots are consistent with the
previous observation of large level spacings at $\tilde{\theta}=0$. At
the beginning (end) of a large level spacings one may expect that
$N_{\mathrm{fluct}}(\tilde{\theta})$ has a statistical tendency to be
positive (negative). At $\tilde{\theta}=0$ we have found exceptionally
large spacings and thus may expect a statistical correlation that a
maximal positive fluctuation occurs before and that a maximal negative
fluctuation (i.e. its minimum) occurs just above $\tilde{\theta}=0$.
This is clearly shown in the plotted integrated densities. In addition
these plots show that these correlations are long ranged and certainly
do not decay on the scale of the mean level spacing $2\pi/N$ (the
scale for spectral $n$-point correlation functions in CUE).

The correlations we have found numerically point to interesting
effects that are currently not understood on a theoretical
level. Analytical approaches to these kinds of correlations would be
highly desirable.

\section{The free energy and its distribution}
\label{sec:free}

\begin{figure}
  \begin{center}
    \includegraphics[width=0.7\textwidth]{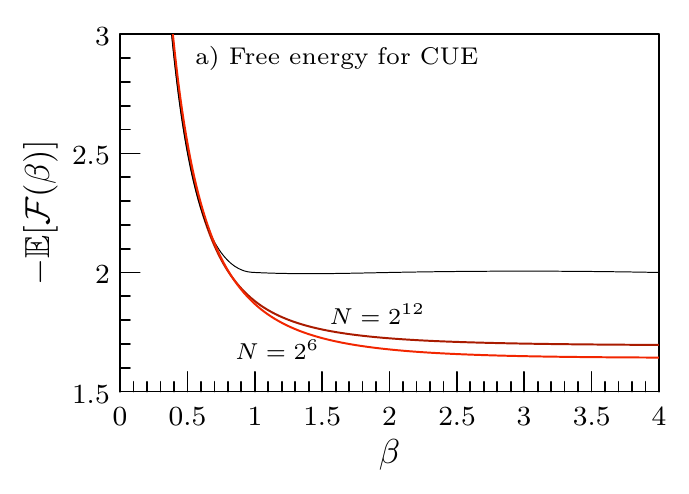}\\
    \includegraphics[width=0.7\textwidth]{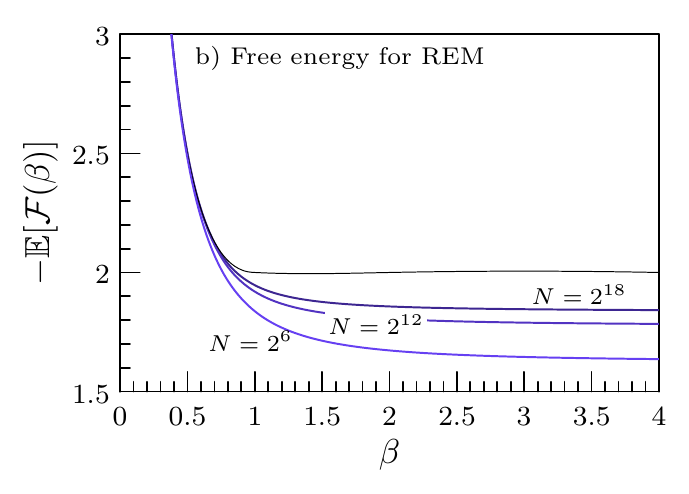}
  \end{center}
  \caption{\label{figure7} (Color online.) 
    Expectation value of the free energy as
    function of the inverse temperature $\beta$. 
    Upper panel: the thick (red) lines give CUE data for 
    $N=2^6$ and $N=2^{12}$ the thin (black) line is the 
    theoretical predicition for $N \to \infty$. 
    Lower panel: the thick (blue) lines give REM data for $N=2^6$, $N=2^{12}$ and $N=2^{18}$ and the thin (black) line is the theoretical predicition for
    $N \to \infty$.\newline Note that the
    scale for the ordinate starts at the value $1.5$.  }
\end{figure}

\begin{figure}
  \begin{center}
    \includegraphics[width=0.7\textwidth]{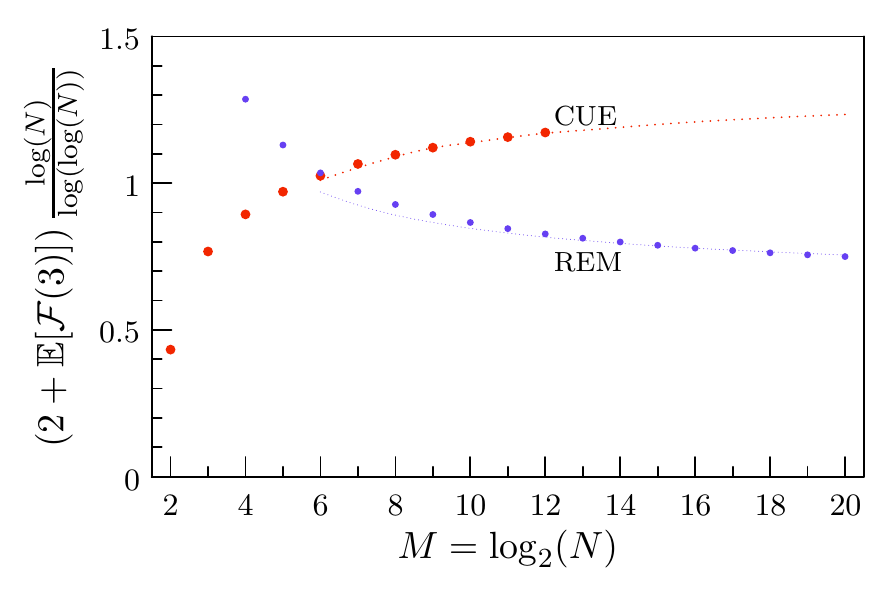}
    \caption{\label{figure8} (Color online.) Convergence of the expected free energy
      at $\beta=3$. The plot shows the difference of the expected
      value $-\mathbb{E}[\mathcal{F}(3)]=2$ and the data for CUE (thick red dots) and REM (thin blue dots), rescaled by
      a factor $\log N /\log(\log N)$. The two thin dotted lines correspond to fitted curves of the form \eqref{eq:fbeta_fit} (see main text for explanation).}
  \end{center}
\end{figure}

In Sec.~\ref{sec:setting} we introduced the partition sum
$\mathcal{Z}_N(\beta)$ and the free energy $\mathcal{F}_N(\beta)$ for
CUE and REM. Numerically they are straightforwardly obtained from the
eigenvalue spectra that we have obtained for both models (by numerical
integration over the spectral angle $\theta$).

In Fig.~\ref{figure7} we plot the expected free energy
$\mathbb{E}[\mathcal{F}_N(\beta)]$ as a function of the inverse
temperature $\beta$ for CUE and REM and some values of $N$. In both
cases we see freezing but the convergence to the predicted curve for
$N \to \infty$ above the freezing transition $\beta\ge 1$ is quite
slow.  In the freezing regime the free energy is dominated by the
maximal value of the modulus of the characteristic polynomial.  In the
previous section we have confirmed the prediction that the latter obey
$2 \log |p_N|_{\mathrm{max}}^2 \sim 2 \log N - c \log \log N + o(1)$
where $c=3/2$ for CUE and $c=1/2$ for REM (see \eqref{eq:scalparam_c}).
For the free energy this implies convergence at a slow rate $\log\log
N/\log N$ in the freezing regime. 
We test this estimate numerically
by considering the difference of the numerically obtained free energy
at $\beta=3$ to the theoretical value $\mathbb{E}[\mathcal{F}(3)]=-2$.
We compare this to a fitted curve of the form
\begin{equation}
  -\mathcal{F}_{\mathrm{fit}}= 2 - c \frac{\log\log N}{\log N} + 
  \frac{g_{\mathrm{fit}}}{\log N}
\label{eq:fbeta_fit}
\end{equation}
where the additional parameter $g_{\mathrm{fit}}$ is fitted to the data at
$N=2^{12}$ for CUE where $g_{\mathrm{fit}}\approx 0.69$ and at $N=2^{20}$ for REM
where $g_{\mathrm{fit}}\approx -0.67$. 
In Fig.~\ref{figure8} we plot the difference of the conjectured 
expectation value $-\mathbb{E}[\mathcal{F}(3)]\to 2$ for $N \to \infty$
and the numerical expectation value at finite values of $N$. 
In the plots we have rescaled the difference by a factor
$\log N/\log(\log N)$). 
For both models the plots are consistent with a
saturation at the value $c=3/2$ or $c=1/2$ for CUE and REM
with higher order deviations $\approx \frac{g_{\mathrm{fit}}}{\log\log N}$.

\begin{figure}
  \begin{center}
    \includegraphics[width=0.7\textwidth]{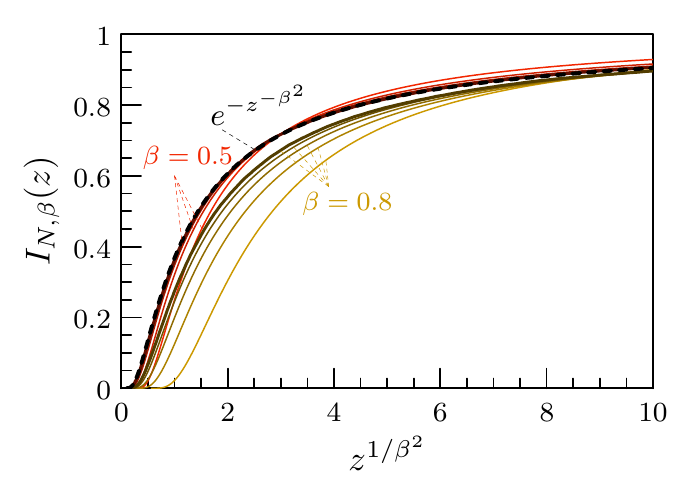}
  \end{center}
  \caption{\label{figure9} (Color online.) Integrated distribution of the rescaled
    partition function for inverse temperatures below the freezing
    transition at $\beta=0.5$ (red curves) and $\beta=0.8$ (orange
    curves).  We plot the curves as a function of $z^{1/\beta^2}$
    where $z$ is the appropriately rescaled partition function (see
    Eq.~\eqref{zscal_subcrit}).  This makes the predicted limiting
    distribution independent of $\beta$.  The different plots are for
    $N=16$, $64$, $256$, $1024$, and $4096$ (with increasing strength
    of color).  The dashed black curve is the predicted curve for 
    $N \to \infty$.
  }
\end{figure}

Next we consider the distribution of the partition function
$\mathcal{Z}_N(\beta)$ for the CUE model (we do not show that data for the
well-understood REM model, because it does not give additional insight
to what we have learned from the comparison so far).  For inverse
temperatures below the freezing transition $\beta<1$ an analytical
prediction is available for the complete distribution.  For this
purpose one rescales the partition function
\begin{equation}
  z= \mathcal{Z}_{N}(\beta)/\mathcal{Z}_e(\beta)
  \label{zscal_subcrit}
\end{equation}
where $\mathcal{Z}_e(\beta)= N^{1+\beta^2}\frac{G^2(1+\beta)}{G(1+2\beta)
  G(1-\beta^2)}$ and $G(x)$ is the Barnes $G$-function.  In the limit
$N \to \infty$ the prediction for the integrated probability
distribution of the rescaled partition function is
\begin{equation}
  I_{\beta}(z)= e^{-z^{-1/\beta^2}}\ .
\end{equation}
In Fig.~\ref{figure9} we plot the numerically obtained integrated
distribution function $I_{N,\beta}(z)$ for CUE for $\beta=0.5$ and
$\beta=0.8$.  For $\beta=0.5$ one sees clear and quick convergence to
the predicted curve.  For $\beta=0.8$, somewhat closer to the freezing
transition, convergence is slower but consistent with the prediction.

In the freezing regime $\beta>1$ a theoretical
prediction for the distribution as $N \to \infty$ 
of the partition
function is known only in the form of a Laplace 
transform (see eq (33) in \cite{FB2008}) to the leading order
\begin{equation}
  \int_0^\infty I_{N,\beta}(Z) e^{-s Z} dZ =
  2 \nu s^{\frac{1}{2\beta} -1} K_1(2 \nu s^{\frac{1}{2\beta}})\ .
  \label{eq:laplace}
\end{equation} 
Here the parameter
\begin{equation}
  \nu= \frac{N^2}{\log^{3/2} N} \nu_{\mathrm{red}}
\end{equation}
sets the overall scale and $\nu_\mathrm{red}$ 
(the \emph{reduced} parameter) is of order $O(1)$.
This implies that the appropriately rescaled partition function 
\begin{equation}
  \tilde{z} = \mathcal{Z}_N(\beta)\frac{\log^{3\beta /2} N}{N^{2\beta}} 
  \label{zscal_supercrit}
\end{equation}
has a finite limiting distribution $I_\beta(\tilde{z})$.
The Laplace transform of the latter is obtained from \eqref{eq:laplace}
by replacing $\nu \mapsto \nu_{\mathrm{red}}$, i.e.
$\int_0^\infty I_{\beta}(\tilde{z}) e^{-s \tilde{z}} d\tilde{z} =
  2 \nu_{\mathrm{red}} s^{\frac{1}{2\beta} -1} 
  K_1(2 \nu_{\mathrm{red}} s^{\frac{1}{2\beta}})$.
The value of the scale $\nu_{\mathrm{red}}$ is not known theoretically
and it is not straightforward to extract numerically from data at finite $N$
because of the generally slow convergence of the model and the fact that
the distribution at finite $N$ depends not only on the scale
factor $\nu$ but also on the shape, which is currently 
not known theoretically. As a practical way to obtain a value
for $\nu_{\mathrm{red}}$ we consider the theoretical expectation value 
of $\log \tilde{z}$ as a function of $\beta$ and $\nu_{\mathrm{red}}$.
Numerically the simple scaling implies that one can extract the full
dependence of $\mathbb{E}[\log \tilde{z}]$ on the parameter $\nu_{\mathrm{red}}$ 
for any given value of $\beta$ by obtaining the inverse Laplace transform
at $\nu_{\mathrm{red}}=1$. At $\beta=3$ the scaling just gives
\begin{equation}
  \mathbb{E}[\log \tilde{z}]\approx 6 \log(\nu_{\mathrm{red}})   + 0.87. 
\end{equation}
We compare this to the numerical fit \eqref{eq:fbeta_fit}
to the data for the free energy
shown in Fig.~\ref{figure8} at the same value $\beta =3$
\begin{equation}
  \mathbb{E}[\log \tilde{z}]= 3 g_{\mathrm{fit}} \approx  2.07
\end{equation}
where the factor $3$ on the right hand side is $\beta$.
This results in an approximate value 
$\nu_{\mathrm{red}} \approx e^{0.2}\approx 1.22$. We do not claim that our 
practical approach gives the correct value as $N \to \infty$. We will
use this scale for comparing the integrated distributions of $\tilde{z}$.

\begin{figure}
  \begin{center}
    \includegraphics[width=0.6\textwidth]{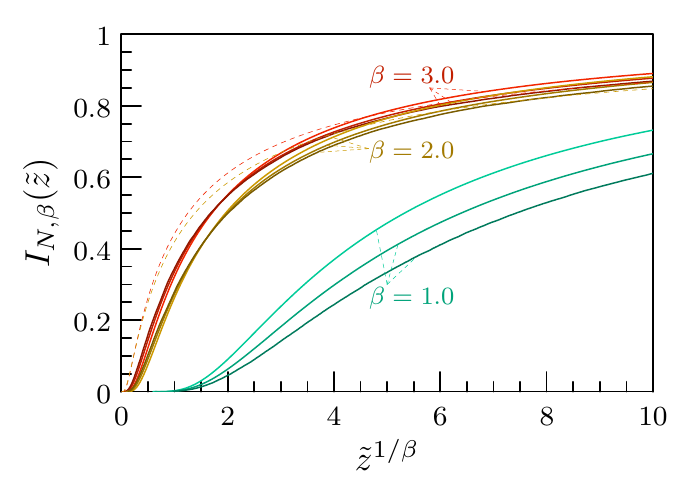}
  \end{center}
  \caption{\label{figure10} (Color online.) 
    Integrated distribution of the rescaled
    partition function for inverse temperatures at the critical value
    $\beta=1$ (green curves) of the freezing transition and in the
    freezing regime at $\beta=2$ (orange curves) and $\beta=3$ (red
    curves).  We plot the curves as a function of
    $\tilde{z}^{1/\beta}$ where $\tilde{z}$ is the appropriately
    rescaled partition function (see Eq.~\eqref{zscal_supercrit}).
    The different plots are for $N=256$, $1024$, and $4096$ (with
    increasing strength of color). The dashed lines give the theoretical
    prediction given by \eqref{eq:laplace} where the overall scale parameter
    $\nu_{\mathrm{red}}$ has been fitted by the procedure outlined in the main 
    text.
  }
\end{figure}

In Fig.~\ref{figure10} we plot the numerically obtained integrated
distributions $I_{N,\beta}(\tilde{z})$ of $\tilde{z}$ at the critical
point ($\beta=1$) and in the freezing regime ($\beta=2$ and
$\beta=3$). The curves in the freezing regime are consistent with the
existence of a limiting distribution on this scale though the convergence to
the predicted limiting distribution may be very slow. Note that the predicted 
curves contain one fitting parameter $\nu_{\mathrm{red}}$. In Fig.~\ref{figure10}
we have used the value $\nu_{\mathrm{red}}=1.22$ obtained from the 
procedure outlined above. Setting $\nu_{\mathrm{red}}=1$ leads to
a limiting curve where the ordinate is stretched by a factor 
$\nu_{\mathrm{red}}^2\approx 1.5$. While this looks much closer to the data 
at the finite values of $N$ we should reiterate that no theoretical
predictions about the leading corrections of the shape of this curve are 
available and one should expect these deviations to decay slowly (e.g. as $1/\log N$ or even slower).
Furthermore note  that our numerics  indicates that at the critical temperature 
$\beta = 1$ the scaling \eqref{zscal_supercrit} is no longer
valid. This is consistent 
with the prediction that a different power of $\log N$  is expected at the
transition 
point (see   eq.(45b) of  \cite{CLD}).

\begin{figure}
  \begin{center}
    \includegraphics[width=0.6\textwidth]{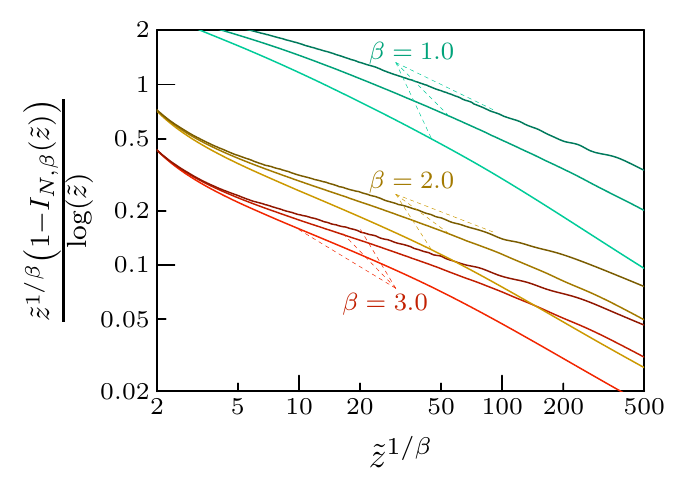}
  \end{center}
  \caption{\label{figure11} (Color online.) 
    Tail of the integrated distribution of the
    rescaled partition function at the critical inverse temperature
    $\beta=1$ (green curves) and in the freezing regime at $\beta=2$
    (orange curves) and $\beta=3$ (red curves).  The tail
    $1-I_{N,\beta}(\tilde{z})$ is divided by $\tilde{z}^{-1/\beta}
    \log \tilde{z}$ according to the predicted behaviour in the
    freezing regime.  The different plots are for $N=256$, $1024$, and
    $4096$ (with increasing strength of colour).  }
\end{figure}

From \eqref{eq:laplace} one may deduce that 
the tails (at large values of the partition function)
behave like
\begin{equation}
  1-I_\beta(\tilde{z}) \propto \tilde{z}^{-1/\beta} \log \tilde{z}
\end{equation}
where the logarithm is related to the strong correlation in CUE.  In
Fig.~\ref{figure11} we plot the ratio of the numerically obtained
tails $1-I_{N,\beta}(\tilde{z})$ and the predicted behaviour for
$\beta>1$.  In order to confirm the prediction one should see the
appearance of a saturation at a finite value as $\tilde{z}$ increases.
While the available numerical data at finite $N$ cannot confirm this
prediction the numerics is consistent with the appearance of such a
saturation as $N \to \infty$.  Indeed the curves for $\beta=2$ and
$\beta=3$ indicate that the behaviour in the tails for finite $N$ may
be of the form $1-I_\beta(\tilde{z}) \propto \tilde{z}^{-1/\beta -
  \nu_\beta(N)} \log \tilde{z}$ where the additional exponent
$\nu_\beta(N)>0$ decays to zero as $N \to \infty$.

\section{Conclusions and Outlook}
\label{sec:conclusions}
The numerical results we have presented here are consistent with the
conjectures put forward in \cite{FHK_prl,FyoKeat2014}.  Taken together
with recent proofs of some of these conjectures, there is growing
evidence supporting the underlying philosophy that the characteristic
polynomials of random matrices behave statistically like
logarithmically-correlated Gaussian fields.  However, the convergence
to the limiting formulae predicted is clearly very slow, and much more
extensive numerical experiments will be required in order to examine
the details of the various conjectures.  The rates of convergence we
have found in our computations are suggestive.  It would be extremely
interesting if they could be verified by a more refined asymptotic
analysis than that carried out in \cite{FHK_prl,FyoKeat2014}.

We believe the correlations we have found between the extreme values
of characteristic polynomials and eigenvalue spacings to be
interesting and worthy of theoretical study.  There has for some time
been a folklore belief that these correlations should exist, but as far as we
understand ours is the first quantitative
study of them.  It may be possible to analyse these by extending the
heuristic asymptotic analysis of \cite{FHK_prl,FyoKeat2014} to mixed
moments involving both the modulus and the argument of the
characteristic polynomial (as in the moment calculations of
\cite{KS2001}).  It should also be possible to do a similar
quantitative analysis for the Riemann zeta-function, where extensive
data exist.

It should be clear that the computations we have described here are
first steps in what we believe may be a worthwhile new line of
research.  We hope that that they will inspire further numerical
studies as well as new theory.

\section{Appendix A: Moments of the partition function}

Let
\begin{equation}
  M_{N,\beta,k}= \mathbb{E}\left[ \mathcal{Z}_N(\beta)^k\right] \
\end{equation}
be the $k$-th moment of the partition function.
A conjecture in  \cite{FHK_prl,FyoKeat2014} states $M_{N,\beta,k} \sim N^{1+\beta^2 k^2}$.
If $\beta$ and $k$ are positive integers one may calculate these
moments directly for arbitrary matrix dimension from the autocorrelation function
of order $k$ for the characteristic polynomials~\cite{Emma}.  It is striking that exact formulae can be 
written down for this quantity, especially at the point of the 
freezing transition.  As a benchmark for our numerics we compare some
low moments with the analytically known values.
Analytically one finds~\cite{Emma}
\begin{align*}
  M_{N,1,1}=& (N+1)N &=& N^2 M_{N,1,1}^{\mathrm{red}}\\
  M_{N,1,2}=&N^2\binom{N+3}{3} &=& \frac{N^5}{6}
                                   M_{N,1,2}^{\mathrm{red}}\\
  M_{N,1,3}=&\frac{N^3(N+5)!}{2520 N!}(N^2+6N+21)&=& \frac{N^{10}}{2520}
                                   M_{N,1,3}^{\mathrm{red}}\\
  M_{N,1,4}=&\frac{8 N^4(N+7)!}{
 13! N!}  \left(7N^6+168N^5+1804N^4\right. && \\
  &\left. \qquad +10944N^3+41893N^2+99624N+154440\right)&=& 
                                                            \frac{56 N^{17}}{13!}
                                                            M_{N,1,4}^{\mathrm{red}}\\
  M_{N,2,1}=&\frac{(N+2)(N+3)!}{12 (N-1)!} &=&\frac{N^{5}}{12}
                                   M_{N,2,1}^{\mathrm{red}}\\
  M_{N,2,2}=&\frac{8N^2 (N+7)!}{15! N!}
              \left(298N^8+9536N^7+134071N^6\right.&&\\
  &\qquad+1081640N^5+549437N^4+18102224N^3&&\\
 & \left. \qquad+38466354N^2+50225040N+32432400\right) &=&\frac{2384N^{17}}{15!}
                                   M_{N,2,2}^{\mathrm{red}} .
\end{align*}
We have here introduced as well the
reduced moments $M^{\mathrm{red}}_{N,\beta,k}$. The reduced moments
obey $M^{\mathrm{red}}_{N,\beta,k} \to 1$ as $N \to \infty$ and are of
order unity for finite values of $N$.  The following tables compare
the known exact values of the reduced moments $M^{\mathrm{red}}_{N,\beta,k}$ with the
corresponding reduced moments
$\overline{\mathcal{Z}_N(\beta)^k}^{\mathrm{\ red}}$ obtained from the numerical data.
In each case the given errors
are one statistical standard deviation. We only present data that has
sufficiently converged (i.e the standard deviation is sufficiently
small). The  statistical errors are consistent with the scaling with
$N$ of the
prediction for higher moments (which determine standard deviations)
$\propto N^{\beta^2 k^2-1/2}$.
\begin{center}
  \begin{tabular}{l|l|l}
    $N$ & $M^{\mathrm{red}}_{N,1,1}$ &
    $\overline{\mathcal{Z}_N(1)}^{\mathrm{\ red}}$\\
    \hline
    $4$ & $1.25$ & $1.25031\pm  0.00013$\\
    $8$ & $1.125$ & $1.12505\pm  0.00018$\\
    $16$ & $1.0625$& $ 1.06230\pm 0.00026$ \\
    $32$ &$1.03125$& $ 1.03145\pm 0.00036$\\
    $64$ &$1.01562\dots$& $ 1.01682\pm 0.00078$\\
    $128$ &$1.0078\dots$& $ 1.0076\pm 0.0013$\\
    $256$ &$1.0039\dots$& $1.0056\pm 0.0023$\\
    $512$ &$1.0019\dots$& $ 1.0014\pm 0.0019$\\
    $1024$ &$1.0009\dots$& $ 0.9959 \pm 0.0046$\\
    $2048$ &$1.0004\dots$& $ 1.011\pm 0.014$\\
    $4096$ &$1.0002\dots$& $ 0.998 \pm 0.055$
  \end{tabular}
\end{center}

\begin{center}
  \begin{tabular}{l|l|l}
    $N$ & $M^{\mathrm{red}}_{N,1,2}$ &
    $\overline{\mathcal{Z}_N(1)^2}^{\mathrm{\ red}}$\\
    \hline
    $4$ & $3.2812\dots$& $3.2835 \pm 0.0010$\\
    $8$ & $1.9335\dots$ & $1.9352\pm 0.0018$\\
    $16$ & $1.4194\dots$ & $1.4158\pm 0.0054$ \\
    $32$ &$1.198\dots$& $1.170\pm 0.012$\\
    $64$ &$1.096\dots$& $ 1.241\pm 0.135$\\
    $128$ &$1.047\dots$& $ 0.856 \pm 0.050$\\
    $256$ &$1.023\dots$& $ 1.238\pm 0.352$\\
    $512$ &$1.011\dots$& $ 0.456\pm 0.029$\\
    $1024$ &$1.005\dots$& $ 0.49 \pm 0.17$
  \end{tabular}
\end{center}

\begin{center}
  \begin{tabular}{l|l|l}
    $N$ & $M^{\mathrm{red}}_{N,1,3}$ &
    $\overline{\mathcal{Z}_N(1)^3}^{\mathrm{\ red}}$\\
    \hline
    $4$ & $56.293\dots$ & $56.362\pm 0.039$\\
    $8$ & $9.794\dots$ & $9.850\pm 0.058$\\
    $16$ & $3.39\dots$ & $3.28\pm 0.14$\\
    $32$ &$1.88\dots$&$ 1.19 \pm 0.10$\\
    $64$ &$1.38\dots$&$ 1.74\pm 0.98$
  \end{tabular}
\end{center}

\begin{center}
  \begin{tabular}{l|l|l}
    $N$ & $M^{\mathrm{red}}_{N,1,4}$ &
    $\overline{\mathcal{Z}_N(1)^4}^{\mathrm{\ red}}$\\
    \hline
    $4$ & $9156.3\dots$ & $9171.0  \pm  12.4$\\
    $8$ & $201.9\dots$ & $ 206.1\pm 5.2$\\
    $16$ &$18.38\dots$  & $1.36 \pm 0.21$
  \end{tabular}
\end{center}

\begin{center}
  \begin{tabular}{l|l|l}
    $N$ & $M^{\mathrm{red}}_{N,2,1}$ &
    $\overline{\mathcal{Z}_N(2)}^{\mathrm{\ red}}$\\
    \hline
    $4$ & $4.9218\dots$ & $4.9255\pm 0.0015$\\
    $8$ & $2.4169\dots$ & $2.4190\pm 0.0023$\\
    $16$ &$1.5968\dots$ & $1.5926\pm 0.0061$\\
    $32$ &$1.273\dots$&$ 1.243\pm 0.013$\\
    $64$ &$1.130\dots$&$ 1.289\pm 0.147$\\
    $128$ &$1.063\dots$&$0.869 \pm 0.056$\\
    $256$ &$1.03\dots$&$1.21\pm 0.36$
  \end{tabular}
\end{center}

\begin{center}
  \begin{tabular}{l|l|l}
    $N$ & $M^{\mathrm{red}}_{N,2,2}$
    &
    $\overline{\mathcal{Z}_N(2)^2}^{\mathrm{\ red}}$\\
    \hline
    $4$ & $27270.2\dots$ & $27313.8\pm  32.9$ \\
    $8$ & $402.9\dots$ & $409.7 \pm 8.4$\\
    $16$ &$ 27.57\dots$ & $21.4\pm 2.9$\\
    $32$ &$5.79\dots$&$ 0.78 \pm 0.18$
  \end{tabular}
\end{center}

\section{Appendix B: On the maximum of characteristic polynomial for
  $\beta-$circular ensemble.}

Consider the circular $\beta-$ ensemble with j.p.d. of real variables
$\theta_i\in [0,2\pi), \, i=1, \ldots, N$ given by \cite{killip}
\footnote{The random matrix Dyson index $\beta>0$ in this Appendix
  should not be confused with the inverse temperature parameter used
  in the main body of the paper.}
\begin{equation}\label{1} {\cal
    P}_{\beta}(\theta_1,\ldots,\theta_N)=\frac{1}{(2\pi)^NC_{N,\beta}}\prod_{i<j}^N|e^{i\theta_i}-e^{i\theta_j}|^{\beta},
  \quad
  C_{N,\beta}=\frac{\Gamma\left(1+N\beta/2\right)}{\Gamma^n(1+\beta/2)},
  \quad \beta>0
\end{equation}
and denote $\left\langle \ldots \right\rangle_{C\beta E_N}$ the
corresponding averages.  Further introduce the characteristic
polynomial
$p_N(\theta)=\prod_{i}^N\left(1-e^{i(\theta_i-\theta)}\right)$ by
Eq.(\ref{defcharpol}) in terms of which we define the partition
function
\begin{equation}\label{3}
  \mathcal{Z}_q=\frac{N}{2\pi}\int_{0}^{2\pi}|p_N(\theta)|^{2q}\,d \theta, \quad q>0
\end{equation}
where $q$ is the inverse temperature (we can not use $\beta$ here for
the inverse temperature as that is reserved for the Dyson index).
Integer moments of the partition function are then given by
\begin{equation}\label{4}
  \mathbb{E}\left\{\mathcal{Z}_q^n\right\}=\frac{N^n}{(2\pi)^n}\int_{0}^{2\pi}\ldots \int_{0}^{2\pi}
  \mathbb{E}\left\{\prod_{l=1}^n |p_N(\theta_l)|^{2q}\right\}\,d \theta_1\ldots d\theta_n
\end{equation}
where
\begin{equation}\label{5}
  \mathbb{E}\left\{\prod_{l=1}^n |p_N(\theta_l)|^{2q}\right\}=\frac{1}{(2\pi)^NC_{N,\beta}}
  \int \prod_{i<j}^N\,|e^{i\theta_i}-e^{i\theta_j}|^{\beta}\prod_{i=1}^N g(\theta_i)\,d\theta_i\equiv \left\langle \prod_{i=1}^N g(\theta_i) \right\rangle_{C\beta E_N}
\end{equation}
and we defined the ``symbol'' function
\begin{equation}\label{6}
  g(\theta)=\prod_{l=1}^n\left(2-2\cos{\left(\theta_l-\theta\right)}\right)^q
\end{equation}
which can be rewritten as
\begin{equation}\label{7}
  \log{g(\theta)}=\sum_{l=1}^n\, q\log\left(2-2\cos{\left(\theta_l-\theta\right)}\right).
\end{equation}
The last equation when compared to Eq.(1.6) from \cite{FF2004} implies
\begin{equation}\label{8}
  a(\theta)=1, \quad  a_l=q, \, b_l=0 \quad \forall l=1,\ldots, n,
\end{equation}
If the parameter $\beta$ is {\it rational} ( that is: $\beta=2s/r$
where $s$ and $r$ are relatively prime) the paper \cite{FF2004} {\it
  conjectured} the generalisation of Fisher-Hartwig formula for $N\to
\infty$, which in the case of (\ref{8}) reads (see (3.13)-(3.14) in
\cite{FF2004} with identification $R=n$ and $q_j=2q/\beta, \, \forall
j)$ as well as $c_n=0, \, \forall n$):
\begin{equation}\label{9}
  \left\langle \prod_{i=1}^N g(\theta_i) \right\rangle_{C\beta E_N, \, N\to \infty}\sim N^{\frac{2nq^2}{\beta}}A^n|_{\left(q_j=2q/\beta\right)} \prod_{l<m}^n\left|e^{i\theta_l}-e^{i\theta_m}\right|^{-\frac{4q^2}{\beta}}
\end{equation}
and the function $A|_{q_j}$ is given by a complicated product of
Barnes functions in eq.(3.11) of \cite{FF2004}.  Substituting the
above to (\ref{4}) and defining $\tilde{q}=q\sqrt{\frac{2}{\beta}}$ we
get:
\begin{equation}\label{10}
  \mathbb{E}\left\{\mathcal{Z}_q^n\right\}= N^{\frac{2nq^2}{\beta}}A^n|_{\left(q_j=2q/\beta\right)} \frac{N^n}{(2\pi)^n}\int_{0}^{2\pi}\ldots \int_{0}^{2\pi}\,
  \prod_{l<m}^n\left|e^{i\theta_l}-e^{i\theta_m}\right|^{-2\tilde{q}^2}\,d\theta_1\ldots d\theta_n
\end{equation}
\begin{equation}\label{10a}
  = \mathcal{Z}^n_e(\tilde{q}) \Gamma(1-n\tilde{q}^2)
\end{equation}
where we introduced the ``typical value'' for the partition function
\begin{equation}\label{10b}
  \mathcal{Z}_e(\tilde{q}) =N^{1+\tilde{q}^2} A|_{\left(q_j=2q/\beta\right)}\frac{1}{\Gamma(1-\tilde{q}^2)}
\end{equation}
We see that the ``freezing temperature'' is now given by the condition
$\tilde{q}=1$ so that $q=\sqrt{\frac{\beta}{2}}$.  The ``free energy''
is given by, to the leading order for $N\to \infty$:
\begin{equation}\label{11}
  -{\cal F}=\frac{1}{q\log{N}} \log{\mathcal{Z}_q}\to \left\{\begin{array}{cc}\frac{1}{q}+\frac{2}{\beta}q, & q<\sqrt{\frac{\beta}{2}}\\
      2\sqrt{\frac{2}{\beta}}, & q>\sqrt{\frac{\beta}{2}}\end{array}\right.
\end{equation}
Correspondingly, the leading order of the maximum of the
characteristic polynomial is given by $
\sqrt{\frac{2}{\beta}}\log{N}$.  To find the subleading order one can
follow the same procedure as for $\beta=2$ and find that the typical
measure of ``high points'' , that is those points where
$2\log{|p_n(\theta)|}>2x\log{N}$ is given by
\begin{equation}
  \mu_e(x)\sim \frac{N^{-\beta x^2/2}}{\log N}\frac{1}{\Gamma\left(1-\beta x^2/2\right)}
\end{equation}
and equating this to $N^{-1}$ we find the "threshold of high values"
to be
\begin{equation}
  x=\sqrt{\frac{2}{\beta}}\left(1-\frac{3}{4}\frac{\log{\log{N}}}{\log{N}}\right)
\end{equation}
which implies two first terms of the maximum to be
$\sqrt{\frac{2}{\beta}}\left(\log{N}-\frac{3}{4}\log{\log{N}}\right)$,
in full agreement with \cite{CMN}.  Finally, the correction term of
the order of unity will be obviously given by the same distribution as
for $\beta=2$, as moments of the partition function are the same, up
to a trivial rescaling by $\sqrt{\frac{2}{\beta}}$ whenever necessary.

We therefore come to the conclusion that as $N\to \infty$ the
characteristic polynomials of circular ensemble for any $\beta>0$ is
essentially described by the same random Gaussian logarithmically
correlated process rescaled by the parameter $\sqrt{\frac{2}{\beta}}$.

\end{document}